\documentclass{aa}
\input psfig.sty
\begin{document}

   \thesaurus{011     
              (11.01.02;  
               11.17.4;  
               13.25.2  
	)	
		}
   \title{On the peculiar X--ray properties of the bright nearby radio-quiet quasar PDS~456.}


   \author{C. Vignali
          \inst{1,2}
          \and
          A. Comastri\inst{2}  
	  \and
          F. Nicastro\inst{3,4}
          \and 
	  G. Matt\inst{5}
 	  \and
	  F. Fiore\inst{3,4,6}
	  \and 
	  G.~G.~C. Palumbo\inst{1}
          }

   \offprints{C. Vignali}

   \institute{Dipartimento di Astronomia, Universit\`a di Bologna, 
	      Via Ranzani 1, I--40127 Bologna, Italy \\
              email: vignali@kennet.bo.astro.it
          \and
	      	Osservatorio Astronomico di Bologna, 
		Via Ranzani 1, I--40127 Bologna, Italy \\
	  \and
                Harvard-Smithsonian Center of Astrophysics,
                60 Garden Street, Cambridge MA 02138,USA \\
          \and
                Osservatorio Astronomico di Roma,
                Via Frascati 33, I--00044 Monteporzio, Italy \\
          \and
		Dipartimento di Fisica, Universit\`a degli Studi ``Roma Tre'', 
		Via della Vasca Navale 84, I--00146 Roma, Italy \\
	  \and
		BeppoSAX Science Data Center, 
		Via Corcolle 19, I--00131 Roma, Italy \\
             }

   \date{Received ???; accepted ??}

   \maketitle
\titlerunning{BeppoSAX and ASCA observations of PDS~456}
\authorrunning{C. Vignali et a.}

   \begin{abstract}

BeppoSAX and ASCA observations of the nearby (z = 0.184), high--luminosity,
radio-quiet quasar PDS~456 are presented.
The X--ray spectrum is characterized by a prominent ionized edge at 8--9 keV
(originally discovered by RXTE, Reeves et al. 2000)
and by a soft excess below 1.5 keV. 
The lack of any significant iron K$\alpha$ emission line suggests for the
edge an origin from line--of--sight material rather than from reflection
from a highly ionized accretion disc. The hard X--ray continuum is indeed 
well modelled by transmission through a
highly-ionized medium with a large column density ($N_{H warm} \sim$ 4.5 $\times$ 10$^{24}$
cm$^{-2}$) plus an additional cold absorber with a lower column density
($N_{H cold}$ $\sim$ 2.7 $\times$ 10$^{22}$ cm$^{-2}$).

   \end{abstract}

%

\section{Introduction}
The average 2--10 keV X--ray spectrum of low/medium-redshift quasars 
observed with EXOSAT (Comastri et al. 1992; Lawson et al. 1992), 
Ginga (Williams et al. 1992; Lawson \& Turner 1997) and ASCA 
(Reeves et al. 1997; George et al. 2000) 
is well represented by a power law with a
slope similar to that of lower luminosity Seyfert 1
galaxies ($\Gamma \simeq$ 1.9--2.0, Nandra \& Pounds 1994) 
and significant dispersion
around the mean value $\sigma \simeq$ 0.2--0.3.
The typical imprints of reprocessing gas (i.e. the reflection ``hump'' 
and the K$\alpha$ emission line), which characterize the spectra
of Seyfert 1 galaxies (Nandra \& Pounds 1994; Nandra et al. 1997) 
are however not present in high luminosity objects. \\
Indeed, only a small number of quasars have a spectrum 
more complex than a simple 
power law (plus, sometimes, a soft excess at low energies). 
This fact has been tentatively ascribed to different physical 
conditions of the accreting gas in the innermost regions surrounding 
high-luminosity quasars with respect to Seyfert galaxies. 
Iwasawa \& Taniguchi (1993) suggested the existence 
(based on Ginga data) 
of an X--ray ``Baldwin'' effect, whereby the equivalent width of the 
iron K$\alpha$ line decreases with luminosity. Nandra et al. (1997, 
1999) have recently confirmed and extended this result using ASCA data. 
Both the profile and strenght of the K$\alpha$ line change with luminosity. 
In particular, for luminosities of 10$^{45-46}$ erg s$^{-1}$, 
ionized iron lines have been detected in a few radio-quiet quasars (RQQs) 
(Nandra et al. 1996; Yamashita et al. 1997; George et al. 2000). 
At higher luminosities there is no evidence for any emission line at 
all (Nandra et al. 1999; Vignali et al. 1999). 
Unfortunately, high-luminosity quasars are usually found at relatively 
high-redshift and therefore are rather weak in X--rays, making the study 
of spectral features extremely difficult. \\
High luminosity quasars differs from Seyfert 1s also in the observed 
shape of the X--ray continuum. 
Recent studies have suggested that high-redshift RQQs (Vignali 
et al. 1999) have flatter X--ray spectral slopes than lower redshift 
objects (George et al. 2000). 
However, it is not clear whether the spectral flattening is 
a function of redshift, implying a flattening of the continuum towards 
high energies, or rather it depends on the luminosity, thus suggesting 
a different emission mechanism at high luminosities. \\
Luminous radio-quiet quasars in the local Universe are rare. In this 
regard, PDS~456 is an exception and could provide hints on the nature 
of some of the properties described above, which are of great relevance 
to the physics of AGNs. PDS~456 is a bright (B=14.7) nearby (z = 0.184, 
Torres et al. 1997) radio-quiet quasar (S$_{1.4~GHz}$ = 22.7 mJy, 
Condon et al. 1998; radio-loudness R$_{\rm L}$ = $-$0.7, Reeves et al. 2000) 
close to the Galactic plane (l = 10$^{\circ}$.39, b=11$^{\circ}$.16; 
N$_{\rm H}$ $\simeq$ 2 $\times$ 10$^{21}$ cm$^{-2}$, Dickey \& Lockman 1990). 
The bolometric luminosity (L $\sim$ 10$^{47}$ erg s$^{-1}$; Reeves et al. 
2000) makes PDS~456 more luminous than the nearby (z = 0.158) 
radio-loud quasar 3C~273, without being jet-dominated. 
The ASCA and RXTE quasi-simultaneous observations discussed by 
Reeves et al. (2000) reveal the presence of 
a deep edge--like feature at E $>$ 8 keV which is likely to
be due to highly ionized iron atoms. The 2--10 keV flux doubled 
its intensity during a strong outburst which lasted for about 4 hours.

In order to further investigate the peculiar properties 
of this quasar we have carried out a medium--deep 
observation of PDS 456 with BeppoSAX.
The results of the analysis are presented in $\S$ 2, compared
with a re-analysis of the ASCA observation in $\S$ 3 and discussed 
in $\S$ 4. 
Throughout the paper a Friedmann cosmology with H$_{0}$ = 50 km 
s$^{-1}$ Mpc$^{-1}$ and q$_{0}$ = 0.0 is assumed.

\section{BeppoSAX observation}

\subsection{Data reduction}
The Italian-Dutch satellite BeppoSAX (Boella et al. 1997a) carries 
four co-aligned Narrow-Field Instruments (hereafter NFI), two of which 
are gas scintillation proportional counters with imaging capabilities: 
the Low Energy Concentrator Spectrometer (LECS, 0.1--10 keV, Parmar et 
al. 1997) and the Medium Energy Concentrator Spectrometer (MECS, 1.5--10 
keV, Boella et al. 1997b). The remaining two instruments are the High 
Pressure Gas Scintillation Proportional Counter (HPGSPC, 4--120 keV, 
Manzo et al. 1997) and the Phoswich Detector System (PDS, 13--200 keV, 
Frontera et al. 1997). 
The quasar was observed by BeppoSAX on 1998 August 19-20. 
Data presented in this paper were obtained only with the imaging 
spectrometers, as PDS~456 has not been detected in the other two instruments.
Standard reduction techniques and screening criteria were applied in 
order to produce useful linearized and equalized event files. 
A total of 62.7 ks and 29.7 ks have been accumulated for MECS and 
LECS instruments, respectively. 
Spectra have been extracted from circular regions of radius 4$\arcmin$ 
and 6$\arcmin$ around the source centroid for the MECS and LECS, 
respectively. 
Background spectra have been extracted from both blank-sky event files 
and from source-free regions in the target field-of-view. No apparent 
difference between the two spectra have been revealed. 
The background contributes to about 10 \% and 25 \% to the MECS and 
LECS count rates, the total net counts being 6.35$\pm{0.10}$ $\times$ 
10$^{-2}$ in the 1.5--10 keV band and 2.12$\pm{0.09}$ $\times$ 10$^{-2}$ 
in the 0.4--4 keV energy range, respectively. \\
The {\sc XSPEC} (version 10, Arnaud 1996) program has been extensively used 
to perform spectral analysis. 
In this paper errors are quoted at 90 \% confidence level for one 
interesting parameter ($\Delta\chi^{2}$ = 2.71, Avni 1976), and energies 
are reported in the source rest frame. 
In all our models, for both neutral and ionized absorbers, we assume solar 
abundances as tabulated in Anders \& Grevesse (1989). 

\subsection{BeppoSAX spectral results}
PDS~456 lies fairly close to the Galactic plane (b $\sim$ 11$^{\circ}$). 
The Galactic H~I column density derived by radio measurements is 
about 2--2.4 $\times$ 10$^{21}$ cm$^{-2}$ 
(Dickey \& Lockman 1990; Stark et al. 1992). 
We have checked whether molecular hydrogen, traced by CO emission,
is present towards the PDS~456 direction, with negative result
(Lebrun \& Huang 1984; Dame et al. 1987). 
Therefore a value of 3 $\times$ 10$^{21}$ cm$^{-2}$ (which is similar 
to the value found from the near-infrared observation of 
PDS~456 by Simpson et al. 1999) for the Galactic column density has 
been employed in all the spectral fits. \\
The combined BeppoSAX MECS and LECS spectra were first fitted with a 
single power law plus Galactic absorption. 
The complexity of the spectrum is evident from the quality of the fit 
($\chi^2$/dof = 253/105) and from the data-to-model ratio (Fig.~1). 
This shows systematic deviations over the entire LECS+MECS band. 
In particular, an excess of counts below $\sim$ 1 keV, as well as a 
strong deficit above 7--8 keV are clearly present, suggesting the 
presence of additional components at those energies. 
Furthermore, the monotonic rising of the model/counts ratio between $\sim 1.5$ and 
$\sim$ 5 keV suggests that additional absorption, 
exceeding the Galactic value, is required. 

\begin{table*}
\caption[]{Results of BeppoSAX LECS (0.4--4 keV) and MECS (1.5--10 keV) 
spectral fits}
\scriptsize
\begin{tabular}{lcccccccc}
\noalign{\smallskip}
\hline
\noalign{\smallskip}
{\bf Model} & $\Gamma_{\rm soft}$ & $N_{\rm H}$ & $N_{\rm H~warm}$ & 
$\Gamma_{\rm hard}$ & $\xi^{a}$/LogU$^{b}$ & $E_{\rm edge}$ & 
$\tau$ & $\chi^{2}$/dof \\
\noalign{\smallskip}
 & & (10$^{22}$ cm$^{-2}$) & (10$^{24}$ cm$^{-2}$) & & & (keV) & & \\
\noalign{\smallskip}
\hline \hline
\noalign{\smallskip}

{\bf A} & \dots & \dots & \dots & 0.94$\pm{0.07}$ & \dots & 
8.77$\pm{0.16}$ & 1.23$^{+0.26}_{-0.23}$ & 143/103 \\

{\bf B} & 3.95$^{+0.78}_{-0.82}$ & 2.40$^{+1.40}_{-0.98}$ & \dots & 
1.29$^{+0.18}_{-0.17}$ & \dots & 8.82$^{+0.21}_{-0.19}$ & 
0.98$^{+0.28}_{-0.26}$ & 97/100 \\ 

{\bf C} & 3.99$^{+0.76}_{-0.80}$ & 2.77$^{+1.29}_{-0.90}$ & \dots & 
1.61$^{+0.20}_{-0.16}$ & 4230$^{+14950}_{-2820}$ & \dots & \dots 
& 94.2/100 \\

{\bf D} & 3.95$\pm{0.79}$ & 2.67$^{+1.34}_{-1.10}$ & 4.47$^{+2.77}_{-3.02}$ 
& 1.44$\pm{0.17}$ & 3.90$^{+0.62}_{-0.56}$ & \dots & \dots & 
94.7/100 \\

\noalign{\smallskip}
\hline

\end{tabular}
\par\noindent
$^{a}$ In units of erg cm s$^{-1}$ \\
$^{b}$ U = n$_{phot}$/n$_e$ is the ionization parameter, defined as 
the ratio of the ionizing photons density at the surface of the cloud 
(n$_{phot}$ = Q/(4$\pi$R$^{2}$c)) and the electron density of the gas. 
\end{table*}

\begin{figure}
\psfig{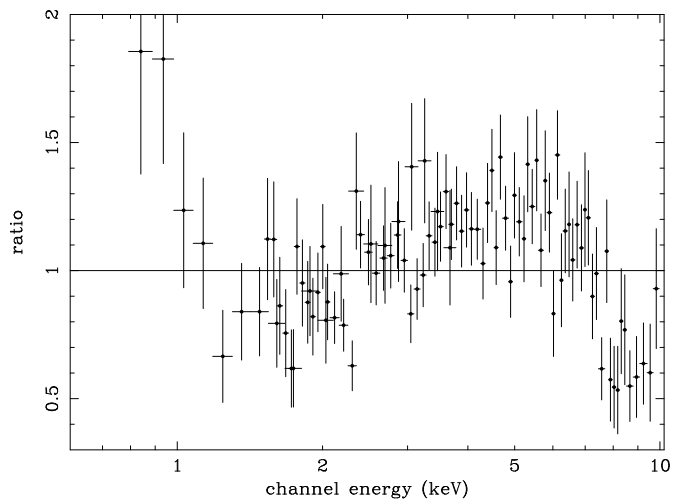}
\caption[]{The data/model residuals for a single power
law fit to the LECS and MECS data.}
\label{fig1}
\end{figure}

To cure the strong residuals at energies above 7 keV (see Fig.~1) we 
added an absorption edge to the power law model (model {\bf A} in 
Table~1). 
The improvement in the fit is highly significant ($\Delta\chi^{2}$/ 
$\Delta$dof = 110/2). 
The best-fit energy of the edge is consistent at 90 \% with that of 
the photoelectric K-edge of \ion{Fe}{xxiv}--\ion{Fe}{xxvi}. If this 
identification is correct, then the measured optical depth (see Table~1) 
implies that a high column density 
($> 10^{24}$ cm$^{-2}$) of highly ionized matter is reprocessing 
the primary X--ray continuum of PDS~456 along our line of sight. 

Though the fit is largely improved by the addition of the absorption 
edge, the $\chi^{2}$ is still unacceptably high (143/103) and the 
residuals continue to show deviations at E $<$ 1 keV. Moreover, 
the spectral index is very flat ($\Gamma$ = 0.94$\pm{0.07}$). 
An acceptable fit and a more reasonable value 
for the power law slope ($\Gamma$ $\simeq$ 1.3) is obtained adding 
to model A two additional spectral components (model {\bf B}): (a) an 
intrinsic neutral absorber obscuring the hard nuclear power law continuum, 
and (b) a soft power law, absorbed by the Galactic column density only. 
The best-fitting soft power law is very steep (though only poorly 
constrained: Table 1), indicating the presence of a luminous 
source of soft photons (L$_{0.5-2 \ keV}$ $\simeq$ 5.2 $\times$ 10$^{44}$ erg s$^{-1}$) 
in the nuclear environment of PDS~456. 
The column density of neutral gas absorbing 
the hard X--ray power law largely exceedes the Galactic value 
along the line of sight to PDS~456, suggesting a ``type-2-like'' 
orientation for this high luminosity radio-quiet quasar. 

\medskip
The phenomenological model B provides a good fit to the 
observed 0.4--10 keV spectrum (Fig.~2) and good constraints 
to the various spectral parameters (Table 1, Fig.~3 and Fig.~4). 
To test the uniqueness of this model and to investigate whether the 
observed soft excess of photons, compared to the absorbed nuclear 
power law, is indeed due to genuine emission and not instead to an 
artifact due to additional complexity in the geometrical or physical 
status of the matter covering our line of sight to PDS~456, 
a few different spectral models were tested. 
More specifically, 
the soft X--ray power law was eliminated and the full-covering 
neutral absorber was either replaced by 
a partial covering absorber, allowing a fraction 
of the nuclear radiation to escape unabsorbed, or by a mildly ionized 
absorber transparent at energies lower than $\sim 0.7$ keV. 
None of these models was able to provide good quality fits to our data. 
We also replaced the soft power law with thermal emission from 
a collisionally ionized plasma, and refitted the data. This did not 
modify the best-fit parameters of the remaining 
spectral components, nor improved further the quality of the fit, 
compared to model B. 
Moreover, a significant contribution to the soft X--rays from 
starburst emission is unlikely. Indeed, given a 60--100 $\mu$m luminosity of 6--10 $\times$ 10$^{45}$ 
erg cm$^{-2}$ s$^{-1}$ (depending on the assumed IR spectral slope) and 
relation [2] in David et al. (1992), only about 5 \% of the soft X--ray emission 
may be due to star-forming emission. 
We then conclude that PDS~456 shows genuine soft X--ray emission 
below $\sim 1$ keV. Such a component is likely to be related to the 
strong UV bump present in the multiwavelength energy distribution 
(Reeves et al. 2000). 
%
\begin{figure}
\psfig{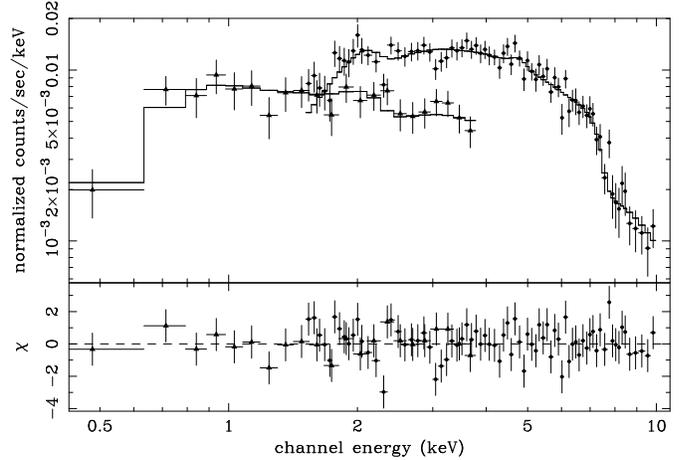}
\caption[]{The BeppoSAX spectrum (MECS$+$LECS, model {\bf B}) and residuals}
\label{fig2}
\end{figure}

\begin{figure}
\psfig{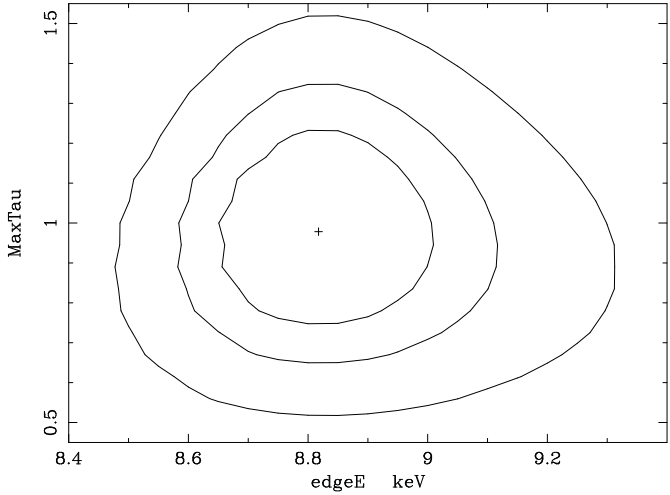}
\caption[]{68, 90 and 99 \% edge energy -- optical depth confidence contours (model {\bf B})} 
\label{fig3}
\end{figure}


\begin{figure}
\psfig{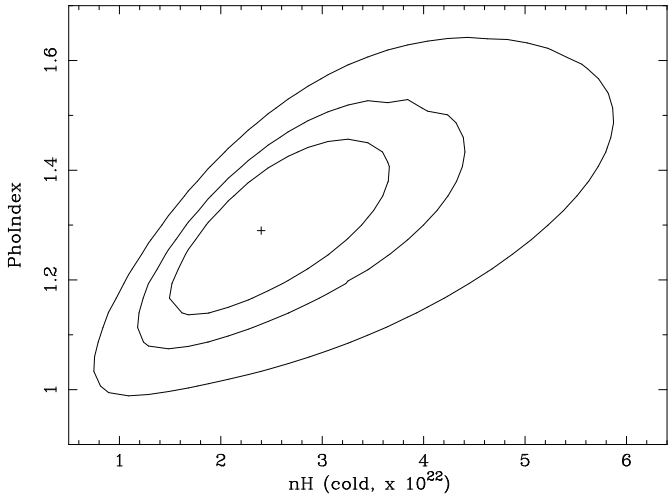}
\caption[]{Intrinsic cold N$_{\rm H}$ -- $\Gamma$ confidence contours for the model {\bf B} in Table~1. 
Absorption in excess to the Galactic value is clearly present }
\label{fig4}
\end{figure}

\medskip
The energy and optical depth of the absorption edge in model B 
strongly suggest the presence of highly ionized gas absorbing and/or 
reflecting the primary radiation. 
To verify this hypothesys we then replaced the photoelectric edge in 
model B, with either an ionized reflector (parameterized by the model 
{\sf PEXRIV} in {\sf XSPEC}, Magdziarz \& Zdziarski 1995: model {\bf C}) or 
an ionized absorber (using the {\sc CLOUDY} - Ferland 1996 - based model 
described in Nicastro et al. 2000a, and adopting the Mathews \& Ferland 1987 
parameterization for the AGN continuum: model {\bf D}), and refitted the data. 
In both cases we left the soft and the hard, reflected/absorbed, power 
law free to vary independently. Both parameterizations gave acceptable 
$\chi^2$. The best-fit values of $\Gamma_{soft}$ and $\Gamma_{hard}$ 
were consistent with each other between the two parameterizations. 
The hard reflected/absorbed continuum ($\Gamma$ $\sim$ 1.4--1.6) is 
still much flatter than the soft X--ray power law. 
Finally, the best-fit ionization parameters are consistent with each 
others, and their values correspond to a very high ionization state of 
the matter reflecting/absorbing the nuclear radiation, with iron mainly 
distributed among species XXI and XXVII. 
Both models provide then an acceptable description of our data. 
However, according to Matt, Fabian \& Ross (1993), such a high ionized 
reflector should produce strong (EW $\sim$ 300--500 eV) fluorescence iron 
K$\alpha$ emission lines at 6.7--6.97 keV. 
These lines are not observed in our data, the 90 \% upper 
limits on their equivalent width being of 120, 115 and 80 eV for 
neutral, He--like and H--like ionic species respectively. 
The corresponding 3$\sigma$ limits lie in the range 150--180 eV.
We then conclude that, while statistically acceptable, the model 
including an ionized reflector is not fully consistent with the 
BeppoSAX data of PDS~456. 
K$\alpha$ emission from highly ionized iron is instead 
not expected to be strong in the case of an ionized absorber. The 
predicted strength of these lines depends on the fraction of solid 
angle covered by the absorber (as seen by the central source), and on 
the dynamics of the gas (see \S 4), but does not exceed few tens of 
eV for the strongest line, fully compatible with our data. 
A spectrum transmitted by a high column ($N_{{\rm H}_{\rm warm}}$ 
$\simeq$ 4.5 $\times$ 10$^{24}$ cm$^{-2}$) of highly ionized matter 
would then self-consistently account for both the observed deep absorption 
FeXXIV-XXVI K edge and the absence of K$\alpha$ line emission from the 
same iron species (model D, Table 1). 

\medskip
The best-fit spectrum gives a 2--10 keV flux of about 5.7 $\times$ 
10$^{-12}$ erg s$^{-1}$ cm$^{-2}$, which corresponds to an intrinsic 
2--10 keV luminosity of about 1.2 $\times$ 10$^{45}$ erg s$^{-1}$. 
At this flux level and given the relatively short exposure time, the source 
has not been detected at high energies by the PDS instrument 
(which spends about half of the time to monitor the background). 
In fact, a 3$\sigma$ detection would have been possible only for a power law 
harder that $\Gamma$ $\simeq$ 1.3 (Guainazzi \& Matteuzzi 1997). 
The present upper limit provides therefore only a loose contraint 
on the high energy spectrum. 

\section{ASCA observation}

\subsection{Data reduction}

PDS~456 was observed by the ASCA satellite (Tanaka et al. 1994) on 
1998 March 7 during the AO6 phase.  The focal plane instruments 
consist of two solid-state imaging spectrometers (SIS, Gendreau 1995) 
and two gas scintillation imaging spectrometers (GIS, Makishima et al. 
1996), characterized by a good spectral resolution (about 2 \% and 
$\sim$ 8 \% at 5.9 keV, respectively) and broad-band ($\sim$ 0.6--10 keV) 
capabilities. 
The observations were performed in FAINT mode and then corrected 
for dark frame error and echo uncertainties as suggested by Otani \& 
Dotani (1994). 
The data were screened with the version 1.4b of the 
{\sc XSELECT} package with standard criteria. 
The final observing time is about 41 ks for both SIS and GIS detectors. 
Source counts were extracted from circles of 6$\arcmin$ radius for GIS 
and 3$\arcmin$.5 for SIS centered on the source, and background spectra 
were extracted from source-free regions from the same CCD chip for SIS 
and from the same field of view for GIS. 
The source count rates (after background-subtraction) are 
6.90$\pm{0.11}$ $\times$ 10$^{-2}$ and 6.13$\pm{0.11}$ $\times$ 10$^{-2}$ 
counts per second for SIS and GIS, respectively. 

\subsection{ASCA spectral results}

SIS and GIS data were fitted  simultaneously allowing the relative 
normalizations to be free to vary, to account for residual 
discrepancies in the absolute flux calibration. 
The residuals of a single power law fit ($\Gamma = 1.48\pm 0.04$, Fig.~5) 
are similar to those of BeppoSAX and fully consistent with the results
reported by Reeves \& Turner (2000) in their analysis of the same ASCA 
data. In particular, a soft excess and an edge appear prominently. 
No emission line has been detected. The 90 \% upper limits for the neutral, 
He--like and H--like lines are in the 45--70 eV range while, the 3$\sigma$
upper limits are of the order of 100--130 eV.

\begin{figure}
\psfig{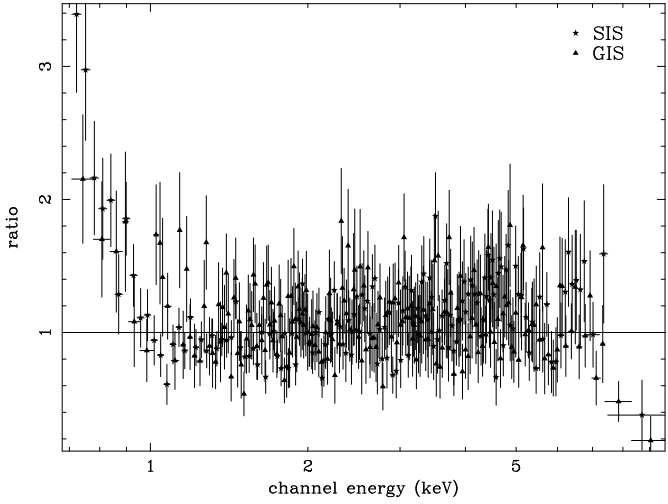}
\caption[]{The residuals obtained by fitting the ASCA spectrum with 
a single power law. The soft excess at E $<$ 1 keV and 
the counts deficit at E $>$ 7 keV are similar to the BeppoSAX data.}
\label{fig5}
\end{figure}

The presence of an ionized edge is fully confirmed by ASCA data 
($\Delta\chi^{2}$/$\Delta$dof = 47/2 with respect to a single power 
law model). 
The edge spectral parameters are consistent with the BeppoSAX observation
but less constrained due to the lower ASCA effective area at 
E $>$ 7 keV.
At low-energies, the spectrum clearly shows a soft component. 
The addition of a power law improves the fit.

Looking for consistency between the BeppoSAX and the ASCA data of 
PDS~456, we then fitted our chosen parameterization of the BeppoSAX data 
(model D) to the ASCA data.  
The edge is fully accounted for by this model. Both the ionization 
parameter and the warm absorbing column density are in agreement with 
BeppoSAX values. 
The ASCA data require a steeper spectral 
slope ($\Gamma_{\bf H}$ $\simeq$ 1.6--2.0) and 
a moderately lower neutral column density, but consistent, 
within the errors, with the BeppoSAX results.\\

\section{Discussion}
ASCA and BeppoSAX observations of PDS~456 were taken about 5 months 
(observer frame) apart. The flux level of the source was lower
during the ASCA observation (F$_{2-10~keV}$ $\simeq$ 3.6 $\times$ 
10$^{-12}$ erg cm$^{-2}$ s$^{-1}$) than during the BeppoSAX one
(F$_{2-10~keV}$ $\simeq$ 5.7 $\times$ 10$^{-12}$ erg cm$^{-2}$ s$^{-1}$).
A 50 \% flux variation over 3 days has been detected by RXTE 
during a quasi--simultaneous observation with ASCA. 
The RXTE range of fluxes encompass the ASCA and BeppoSAX measurements.
In addition, a strong flare with a doubling time of about 17 ksec 
is present in the RXTE observation (Reeves et al. 2000).
The underlying continuum of the ASCA/RXTE spectrum 
($\Gamma \simeq$ 2.3--2.4) is steeper than the slope obtained from the 
present analysis of BeppoSAX and ASCA data and from the independent 
analysis of the ASCA spectrum discussed in Reeves \& Turner (2000).
The origin of such a discrepancy is unclear.
Intercalibration errors between RXTE and ASCA and/or a significant 
steepening of the continuum spectrum above 10 keV may provide a 
plausible explanation.

\medskip
The presence of a prominent ionized iron K edge, with 
similar values for the best-fit energy and optical depth in RXTE 
and BeppoSAX data, represents the most striking observed feature in 
PDS 456 and unambiguosly points to a highly ionized nuclear environment 
in this quasar.
The edge parameters are equally well reproduced by either 
reflection off or transmission through a highly ionized gas.
In the first case the ionization status inferred from the edge 
energy would imply a strong ($\sim$ 300--500 eV; Matt et al. 1993) 
ionized (6.7--6.97 keV) iron line. 
If the edge originates in a high column density (N$_{\rm H}$ a few  
10$^{24}$ cm$^{-2}$) warm gas, no strong iron line is expected (see below).
The BeppoSAX and ASCA upper limits on the line intensity derived 
from the present analysis would then favour a transmission model. On the 
other hand, the tentative detection of a broad iron line in the RXTE data 
($\sigma \sim$ 1 keV, EW $\sim$ 350 eV) would instead favour the 
reflection scenario (Reeves et al. 2000).

To further investigate the properties of the high column 
density, highly ionized absorber we have carried out more detailed 
calculations using the photoionization models described in Nicastro, Fiore 
and Matt (1999) and Nicastro et al. (2000a), which 
includes photoelectric and resonant absorption as well as gas emission. 
The best-fit value of the ionization parameter implies that the
iron atoms are equally distributed among the 3 highest ionization states: 
 \ion{Fe}{xxv} ( $\sim$ 30 \%), \ion{Fe}{xxvi} ($\sim$ 40 \%) and 
\ion{Fe}{xxvii} (i.e. fully ionized, $\sim$ 30 \%). 
In a spherical configuration, the net intensity of the corresponding 
emission lines depends (a) on the fraction of solid angle (as seen 
by the central source) covered by the absorber/emitter, and (b) the 
ratio between the outflowing ($v_{out}$) and the turbolence ($v_{turb}$) 
gas velocities. We then ran our models for different values of the 
covering factor between 0.1 and 1. 
To maximize the net contribute of line emission, compared to absorption, 
we used $v_{out}$/$v_{turb}$ = 5, which guarantees a 
peak-to-peak separation of absorption and emission lines by the same 
transition greater than 3 times the width of these features (Nicastro et al. 2000b). 
The rather uncertain knowledge of the dynamical status of the gas 
prevents us from a more detailed treatment. 
The results indicate that the total equivalent width of the 
line blend (including recombination, fluorescent and intercombination 
lines) never exceeds $\sim$100 eV, a value lower than the upper limits 
derived from ASCA and BeppoSAX data, leading further support to the 
transmission model. 
Finally, the 0.1-2 keV continuum-plus-line fluxes predicted by 
these models allowed us to estimate the maximum contribution of gas 
emission to the soft component measured in both the BeppoSAX and ASCA 
data to be not larger than $\sim$ 20 \%. 

\section{Conclusions}

The most important results obtained from the BeppoSAX observation of 
the luminous quasar PDS~456 can be summarized as follows:
\begin{itemize}

\item[$\bullet$]
The BeppoSAX observation confirms the presence of a deep ionized K 
edge discovered by RXTE at $\sim$ 8.5--9 keV implying the presence
of highly ionized gas around the central source. \\

\item[$\bullet$]
The absence of significant iron line emission in both the
BeppoSAX and ASCA data favours a model where the hard X--ray continuum
is absorbed by a high column density of highly ionized gas.
Further spectral complexity is present at lower energies, requiring the
presence of an additional cold (or weakly ionized) absorber. \\

\item[$\bullet$]
The high luminosity of PDS~456 ($L_{\rm bol}$ $\sim$ 10$^{47}$ erg s$^{-1}$) 
and the properties of the absorbing medium which partially hides 
the X--ray source make this object a good candidate to test the emission 
mechanisms and reprocessing in quasars. \\ 

\end{itemize}

\par\noindent 
Further investigations with {\sc CHANDRA} and {\sc XMM} 
would be extremely helpful to assess the nature of the X--ray emission 
(both in the soft and in hard X--rays domain) and the ionization state 
of the absorbing matter. 
Moreover, X--ray spectroscopy of others bright, relatively nearby quasars 
would be very important to check whether PDS 456 is an exceptional object or
the imprints of highly ionized high column density gas are 
common in luminous quasars. 
A better knowledge of the X--ray properties of bright AGN would also 
allow to probe the physics of accretion processes at high luminosities.

\begin{acknowledgements}
We thank all the people who, at all levels, have made possible the 
BeppoSAX mission. 
We also thank all the members of the ASCA team who operate the satellite and 
maintain the software and database. 
This work has made use of the NASA/IPAC Extragalactic Database (NED) 
which is operated by the Jet Propulsion Laboratory, Caltech, under contract 
with the National Aereonautics and Space Administration, of data obtained 
through the High Energy Astrophysics Science Archive Research Center 
Online Service, provided by the Goddard Space Flight Center and of the 
Simbad database, operated at CDS, Strasbourg, France. 
The authors would like to thank the referee, S. Komossa, for her 
suggestions who improved the quality of the paper. 
Financial support from Italian Space Agency under the contract ARS--98--119 
is ackowledged by C.~V., A.~C and G.G.C.~P. This work was partly supported 
by the Italian Ministry for University and Research (MURST) under grant 
Cofin98-02-32. F.N. Aknowledge the NASA grant NAG5-9216. 
\end{acknowledgements}


\begin{thebibliography}{}

\bibitem[]{} Anders E., Grevesse N., 1989, Geochimica et Cosmochimica 
Acta 53, 197

\bibitem[]{} Arnaud K.~A., 1996, in: Astronomical Data Analysis Software and 
Systems V, Jacoby G., Barnes J. (eds.), ASP Conf. Series, vol.~101, 17

\bibitem[]{} Avni Y., 1976, ApJ 210, 642
%

\bibitem[]{} Boella G., Butler R.~C., Perola G.~C., et al., 1997a, A\&AS 122, 299

\bibitem{d0} Boella G., Chiappetti L., Conti G., et al., 1997b, A\&AS 122, 327

\bibitem[]{} Comastri A., Setti G., Zamorani G., et al., 1992, ApJ 384, 62

\bibitem[]{} Condon J.~J., Cotton W.~D., Greisen E.~W., et al., 1998, AJ 115, 1693

\bibitem[]{} Dame T.~M., Ungerechts H., Cohen R.~S., et al., 1987, ApJ 322, 706

\bibitem[]{} David L.~P., Jones C., Forman W., 1992, ApJ 388, 825

\bibitem[]{} Dickey J.~M., Lockman F.~J., 1990, ARA\&A 28, 215

%
%
%
\bibitem[]{} Ferland G.~J., 1996, CLOUDY: 90.01

\bibitem[]{} Frontera F., Costa E., Piro L., et al., 1997, A\&AS 122, 357


\bibitem[]{} Gendreau K., 1995, PhD Thesis, Massachussetts Institute of Technology

\bibitem[]{} George I.~M., Turner T.~J., Yaqoob T., et al., 2000, ApJ 531, 52

\bibitem[]{} Guainazzi M., Matteuzzi A., 1997, SDC Technical Report

%
%

\bibitem[]{} Iwasawa K., Taniguchi Y., 1993, ApJ 413, L15

\bibitem[]{} Lawson A.~J., Turner M.~J.~L., Williams O.~R., et al., 1992, MNRAS 259, 743

\bibitem[]{} Lawson A.~J., Turner M.~J.~L., 1997, MNRAS 288, 920

%
\bibitem[]{} Lebrun F., Huang Y.-L., 1984, ApJ 281, 634

%
\bibitem[]{} Magdziarz P., Zdziarski A.~A., 1995, MNRAS 273, 837

\bibitem[]{} Makishima K., Tashiro M., Ebisawa K., et al., 1996, PASJ 48, 171

\bibitem[]{} Manzo G., Giarrusso S., Santangelo A., et al., 1997, A\&AS 122, 341

\bibitem[]{} Mathews W.~G., Ferland G.~J., 1987, ApJ 323, 456

\bibitem[]{} Matt G., Fabian A.~C., Ross R.~R., 1993, MNRAS 262, 179

%

\bibitem[]{} Nandra K., Pounds K.~A., 1994, MNRAS 268, 405

\bibitem[]{} Nandra K., George I.~M., Turner T.~J., Fukazawa Y., 1996, ApJ 464, 165

\bibitem[]{} Nandra K., George I.~M., Mushotzky R.~F., Turner T.~J., Yaqoob T., 
1997, ApJ 476, 70

\bibitem[]{} Nandra K., 1999, in "Quasars and Cosmology", 
Eds. G. Ferland and J. Baldwin, Astronomical Society of the Pacific, 
San Francisco, CA. (astroph/9907192)

%

\bibitem[]{} Nicastro F., Fiore F., Matt G., 1999, ApJ 517, 108

\bibitem[]{} Nicastro F., Piro L., De Rosa A., et al., 2000a, ApJ 536, 718

\bibitem[]{} Nicastro F., Elvis M., Fiore F., Matt G., Savaglio S., 
2000b, Proceedings of the Conference ``X--ray Astronomy '99: Stellar Endpoints, 
AGNs and the Diffuse X--ray Background'' 

\bibitem[]{} Otani C., Dotani T., 1994, ASCA Newsl. 2, 25

\bibitem[]{} Parmar A.~N., Martin D.~D.~E., Bavdaz M., et al., 1997, A\&AS 122, 309

\bibitem[]{} Reeves J.~N., Turner M.~J.~L., Ohashi T., Kii T., 1997, MNRAS 292, 468

\bibitem[]{} Reeves J.~N., O'Brien P., Vaughan S., et al., 2000, MNRAS 312, L17

\bibitem[]{} Reeves J.~N., Turner M.~J.~L., 2000, MNRAS 316, 234

%
%

\bibitem[]{} Simpson C., Ward M., O'Brien P., Reeves J., 1999, MNRAS 303, L23

\bibitem[]{} Stark A.~A., Gammie C.~F., Wilson R.~W., et al., 1992, ApJS 79, 77

\bibitem[]{} Tanaka Y., Inoue H., Holt S.~S., 1994, PASJ 46, L37


\bibitem[]{} Torres C.~A.~O., Quast G.~R., Coziol R., et al., 1997, ApJ 488, L19

%
%
\bibitem[]{} Vignali C., Comastri A., Cappi M., et al., 1999, ApJ 516, 590

\bibitem[]{} Williams O.~R., Turner M.~J.~L., Stewart G.~C., et al., 1992, ApJ 389, 157

%
\bibitem[]{} Yamashita A., Matsumoto C., Ishida M, et al., 1997, ApJ 486, 763

\end{thebibliography}
\end{document}